# Profit and loss manipulations by online trading brokers

Golnaz Shahtahmassebi* and Lascelles (Lez) Wright[1]**

*School of Science and Technology, Nottingham Trent University
Contact: golnaz.shahtahmassebi@ntu.ac.uk
**Lez Wright Personal Development, Lez Wright Peak Performance

**Abstract:** Online trading has attracted millions of people around the world. In March 2021, it was reported there were 18 million accounts from just one broker.

Historically, manipulation in financial markets is considered to be fraudulently influencing share, currency pairs or any other indices prices.

This article introduces the idea that online trading platform technical issues can be considered as brokers manipulation to control traders (consumers) profit and loss. More importantly it shows these technical issues are the contributing factors for retail traders' losses of the 82% risk of losing money identified by the FCA.

We identify trading platform technical issues of one of the world's leading online trading providers and calculate retail traders losses caused by these issues. To do this, we independently record each trade details using the REST API response provided by the broker. We show traders log activity files is the only way to assess any suspected profit or loss manipulation by the broker. Therefore, it is essential for any retail trader to have access to their log files.

We compare our findings with broker's Trustpilot customer reviews. We illustrate how traders' profit and loss can be negatively affected by broker's platform technical issues such as not being able to close profitable trades, closing trades with delays, disappearance of trades, disappearance of profit from clients statements, profit and loss discrepancies, stop loss not being triggered, stop loss or limit order triggered too early.

Although regulatory bodies try to ensure "that consumers get a fair deal", these attempts are hugely insufficient in protecting retail traders. Therefore, regulatory bodies such as the FCA should take these technical issues seriously and not rely on brokers' internal investigations, because under any other circumstances, these platform manipulations would be considered as crimes and connivingly misappropriating funds.

---

[1]Authors' disclaimer: It would have been impossible to write this paper without extensive research, actively trading for five years, considerable knowledge of the broker's platform and expert knowledge in REST API.

# 1   Introduction

Online trading has attracted millions of people around the world. In March 2021, from just one broker, Robinhood Markets reported 13 million users (Martin and Wigglesworth 2021). Also 18 million customers' bank accounts were linked to Robinhood Markets in March 2021, an increase from 7.2 million in 2020 (Fitzgerald 2021).

Online foreign exchange (forex) spread betting[2] and day trading with low transactions costs (Paton and Williams 2005) have become a modern and attractive approach for risky investments that can result in a significant return or losing the entire one's fortune (Barnes 2018; Cassidy, Pisac, and Loussouarn 2013; Moeeni and Tayebi 2018; Shange, Duan, and Lui 2021). According to the UK's Financial Conduct Authority (FCA), 82% of customers have always lost money on spread betting and similar market products (Financial Conduct Authority 2016).

However, the emotional distress of losing money can be equivalent to being traumatised, harassed or even the loss of a loved one. In some extreme situations it results in successful/unsuccessful suicidal attempts (Arthur, Williams, and Delfabbro 2016; Duan, and Lui 2021; LaPlante et al. 2009; Sharman et al. 2019).

In some cases, traders' financial losses are due to broker market manipulation (Allen, Litov, and Mei 2006; Langevoort, 1996; Putins 2012; Partington 2018). Regulatory bodies aim to "make financial markets work well so that consumers get a fair deal" (Financial Conduct Authority 2021). However, these attempts are hugely insufficient in protecting retail traders from brokers manipulating their trading platform through technology in ways that it is almost impossible to identify in isolation.

The technology issues on broker online trading platforms will not be seen as manipulation by the regulatory bodies but merely regarded as program bugs, technical glitches (Cardella et al. 2014) or technical issues (Seyfert 2016) when in fact these so-called online trading platform issues are more sinister and nefarious.

In January 2021, traders experienced inability to log in to their account, disappearance of open positions and unable to close positions on IG's trading platform (Saks-McLeod 2021) of which could assist the broker to control traders' profit and loss. In January 2021, IG, Trading 212 and Robinhood Markets banned traders from trading on GameStop and AMC and from opening new positions by locking traders out of their trading platform (Reuters 2021; Leonhardt 2021; Sharma 2021).

In this research article, we illustrate how brokers manipulate traders' profit and loss through so-called online trading platform technical issues. It is impossible to identify these manipulations only based on the transaction information provided by the broker. Therefore, we independently record each trade details using REST API[3] response provided by the broker. This is the only way to detect if a broker is manipulating its online trading platform to control profit and loss of its clients.

---

[2] Spread betting is a way of speculating on the price movement of financial instruments, such as forex and shares (CMC Markets 2021).

[3] "REST stands for REpresentational State Transfer. It means when a RESTful API is called, the server will transfer to the client a representation of the state of the requested resource" (Avraham 2017).

In addition, this paper presents how the online demo trading platform can provide a deceptive and enticing experience to encourage traders to spend real money on the online live trading platforms (Bencino 2009).

As an example, we identify trading platform technical issues of one of the world's leading online trading providers, IG[4], which operates across seventeen countries (IG 2021). We show how traders' profit and loss can be negatively affected by IG's online trading platform technical issues such as not being able to close profitable trades, closing trades with delays, disappearance of trades, disappearance of profit from clients statements, profit and loss discrepancies, stop loss not being triggered, stop loss or limit order triggered too early.

Now that we live in a digital world, customer reviews have become an essential part of our online purchasing activities, especially when making a decision about purchasing a product or services (Karakaya and Barnes 2010). One pays extra attention to negative reviews so that they can make informed decisions. For this reason, we use Trustpilot reviews that can be a true representation on how customers are thinking about a particular business.

Using IG's Trustpilot 1 star reviews (Hu, Bezemer, and Hassan 2018; Levy, Duan, and Boo 2012; Luca 2016; Palomba et al. 2015), our study shows the broker's issues are still ongoing and not isolated to specific time periods. Looking at IG's clients' reviews such issues have existed since 2015 and have had a negative effect on retail traders' financial assets and investments to the point that it resulted in one successful suicide attempt as highlighted by a reviewers' comment.

The remainder of this paper is organised as follows. Section 2 provides a description of the research question, available data and statistical methodology used to analyse the data. Section 3 illustrates the results and Section 4 discusses the implication of these results. Finally, Section 5 provides concluding remarks.

The analysis for this paper was done in R 4.0.3, on, 20.04 on a PC with six core CPU, 32Gb RAM and Nvidia GPU with 4Gb memory. The internet connection has the speed of 1Gbs. The REST API trading requests were sent using Node.js v14.15.4.

## 2 Research question, data and methodology

Using the data described in this section, we attempt to answer the following research question:

*Research question:* Are spread betting brokers manipulating their online trading platform by so called technical issues to control profit and loss of their clients?

This section provides the details of traded currency pairs, methods for collecting information related to IG's technical issues, Trustpilot reviews and statistical methods used for analysing these reviews.

---

[4] IG is a trading name of IG Index Limited.

## 2.1 Currency pairs

During 4 May 2021 - 22 June 2021, 168 trades were made. Only 155 of those trades were recorded on IG's transaction history and 13 of those have disappeared. Table 1 provides the list of currency pairs traded between 4 May 2021 - 22 June 2021 and used in this study.

|  | | Secondary Currency (columns) | | | | | | | |
|---|---|---|---|---|---|---|---|---|---|
|  | | AUD | CAD | CHF | DKK | HUF | JPY | NZD | USD |
| Primary Currency (rows) | AUD/ |  | 4 |  |  |  | 12 |  | 4 |
|  | CAD/ |  |  |  |  |  | 1 |  |  |
|  | CHF/ |  |  |  |  |  | 3 |  |  |
|  | EUR/ | 9 | 6 |  |  |  | 19 | 10 | 3 |
|  | GBP/ | 12 | 6 | 5 |  |  | 20 | 3 | 18 |
|  | NZD/ |  | 6 |  |  |  | 1 |  | 3 |
|  | USD/ |  | 14 |  | 1 | 1 | 7 |  |  |

Table 1: 23 currency pairs traded between May 2021 - June 2021. Rows show the primary currency and columns show secondary currency, e.g., AUD/CAD was traded 4 times or EUR/JPY was traded 19 times during this period.

## 2.2 Trade information

Three separate sets of information were recorded about each position of which all come from IG: transaction history, REST API responses and trader's activity log files.

IG transaction history consisted of date, currency pair, opening position time and price, closing position time and price, size of the trade which -1 shows a sell position and 1 indicating buy position.

REST API response (Figure 1) information recorded by the trader consisted of a unique dealId, currency pair, buy/sell, size of the trade, opening position time and price, closing position time and price, profit/loss, amount of pip that the market has moved since opening a position and any possible error messages if the request was not executed.

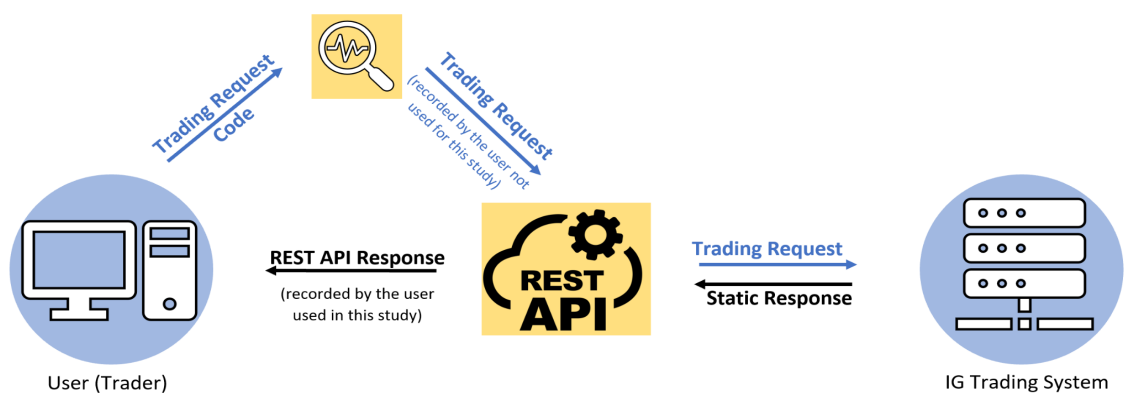

Figure 1: User (trader) REST API trading structure used in this study to connect with IG's trading system and record trading information.

Trader's activity log file recorded on IG's servers which should be the most detailed information. These should show a detailed history of all trader's interaction with IG's online trading platform. It comprises time, type of order, channel (i.e., was it manual trading or REST API trading), Market (currency pair), summary of the activity, result (i.e., was the request successful or not), times of receiving and executing the order, size of the order and finally the price.

All trades were done in British pound (GBP). No details of prices and trades are presented.

## 2.3 Trustpilot Reviews

We used Trustpilot reviews, downloaded on 17 July 2021, to assess the commonality between identified issues in this study and issues that other clients have reported since 2015 (Larsson 2019). Trustpilot reviews when verified can be a true representation on how customers are thinking about a particular business and can empower clients (Constantinides and Holleschovsky 2016; Fawkes and Greggory 2000; Ghazi 2017; Mariani and Nambisan 2021; Mladenovic, Krajina, and Milojevic 2019). To our knowledge, there were no Google reviews available for this selected broker and all Trustpilot reviews were verified.

Out of three Trustpilot review pages about IG, there was only one page that was claimed and verified by IG which was the only one considered for this study. The collected review data consists of review counts, rating, date of publication, date of response, whether the review was verified, whether IG invited customers to review, the content of the review and the reply from IG.

Identification of common technical issues were based on the 1 star comments (Hu, Bezemer, and Hassan 2018; Levy, Duan, and Boo 2012; Palomba et al. 2015; Luca 2016). We could not extract any technical issues from 5 star reviews because 82% of 5 star reviews were from new customers who complimented IG for activating their accounts.

Any ratings without comments were excluded.

## 2.4 Statistical Analysis

The correlation between categorical variables (e.g., rating and review by invitation) was assessed using Pearson's chi-squared test (Plackett 1983).

Each trading platform issue is presented by its frequency, which means the number of times each issue appeared during this study or mentioned in 1 star IG's Trustpilot reviews.

In order to show how profit and loss is controlled by IG, we calculated the total financial loss (in GBP) at the presence of each online trading platform technical issue. The financial loss is defined as the difference between profit and loss recorded from IG's REST API response and IG transaction history. In the case of Trustpilot reviews, the financial loss was considered as the loss (in GBP) claimed by the reviewer.

To investigate the trend of 1 - 5 star ratings between 2015 and 2021, the number of 1 - 5 star ratings in each year were counted and compared with each other. The daily count of 1 star IG's Trustpilot reviews reporting any platform technical issues was recorded and presented year by year from 2015 to 2021. This helps to identify any observable patterns on IG's platform technical issues.

We used the cleanNLP package in R (Arnold 2017) to perform text analysis, annotate feedback and identify the most common issues that traders experienced among Trustpilot reviews (Feinerer and Hornik 2008; Jockers 2015).

## 3 Results

This section provides details of IG's trading platform technical issues, associated financial losses and IG's customer service experience based on recorded trading data and IG's Trustpilot reviews. It also presents how the online demo trading platform can provide deceptive information that encourages traders to trade on the online live trading platform.

### 3.1 Platform Reliability Assessment

Between 4 May 2021 to 22 June 2021, we identified fifteen serious technical issues on IG's online live trading platform. In total, 168 trades from 23 currency pairs were recorded on IG's transaction history. There were 13 trades recorded from IG's REST API responses at the time of closing those trades which have now disappeared from IG's transaction history. Of 155 trades remaining on the transaction history, 90 trades were closed with profit and 65 trades were closed with losses.

The disappeared profit (financial loss) due to disappearance of 13 trades technical issues was equivalent to £1440. In the following, reported financial losses related to other technical issues excludes the 13 disappeared trades because the data was deleted from transaction history.

Table 2 presents the details of each issue, frequency of occurrence and financial losses associated with each issue. It can be seen that disappearance of the trades has the largest effect on recorded financial loss. Profit and loss discrepancy is the second largest contributor to the total amount of the financial loss.

It was noted that the overall profit displayed on the IG transaction history was 50% less than the recorded data from the REST API responses. The loss was 50% more than the recorded data from REST API responses. Which means, when trading, if you win, you win 50% less than you should win, and if you lose, you lose 50% more than you should lose.

The third factor with the largest effect is failure of the online trading platform to close a trade. Out of 155 trades, the closing request for 83 open positions failed. Those positions had to be closed manually and with a maximum of 3 hours delay which resulted in the loss of £371. In general, failure to close trades or delays in closing trades were the most likely technical issue to happen, affecting 61% of trades.

Figure 2 (a) illustrates profit and loss discrepancies grouped by traded currency pairs. It can be seen that GBP pairs have the highest number of discrepancies. Figure 2 (b) shows a scatter plot of the profit and loss (in pips) discrepancies against time of the day and suggests that the largest discrepancy can happen during 5am to 3pm.

| | Frequency (number of times these happened) | Financial Damage |
|---|---|---|
| The REST API closing position request is unreliable and unstable. | 83 | 215 |
| The open position discrepancies between Broker's online trading platform and IG' REST API & log files. | 51 | |
| Delays in opening position requests. Up to twenty seconds | 51 | 113 |
| The close position discrepancies between Broker's online trading platform and IG' REST API & log files. | 39 | |
| The profit presented on Broker transaction history is consistently 50% less. | 16 | 195 |
| Loss presented on Broker transaction history is consistently 50% less. | 16 | 277 |
| Delays in closing position requests. Up to thirteen seconds | 12 | 109 |
| Unreliable & unstable login, not executing REST API closing position requests. | 7 | |
| Executing opening position requests with REST API's confirmed error messages. | 5 | 46 |
| Duplicating opening position requests with REST API's confirmed error messages. | 5 | 57 |
| Open positions with REST API's confirmed error messages must be closed manually. | 5 | |
| Unreliable & unstable login, not executing REST API opening position requests. | 5 | |
| Unstable transaction history | 5 | |
| The REST API opening position request is unreliable and unstable. | 3 | |
| Disappearance of trades placed using REST API. | 1 | 1440 |
| ***Total*** | **304** | **2452** |

Table 2: Summary of the issues identified and the financial damages that they caused.

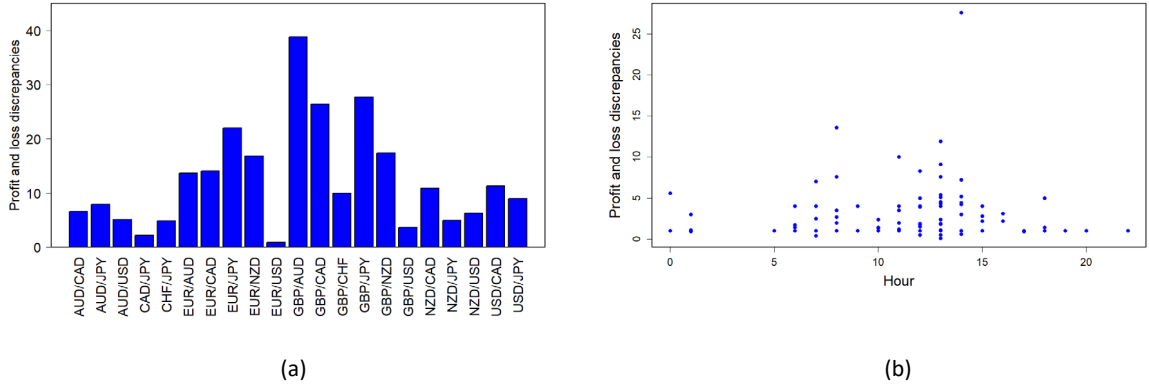

(a)                          (b)

Figure 2: (a) The profit and loss (in pips) discrepancies grouped by currency pairs. (b) Scatter plot of the profit and loss (in pips) discrepancies against time of the day. The discrepancies are more likely to happen between 5am to 3pm.

## 3.2 Demo and Live Platforms

It is common for any trader before joining a new online trading platform or using a new strategy to test the broker's online trading platform with their demo account. Such a demo account should normally provide exact features of the live account with the same live data feed. Therefore, once a test is done to the satisfaction of the trader, the trader will decide whether to move on to the live trading platform or not (Bencino 2009).

Table 3, illustrates the minimum bet size available on online demo trading platforms compared to the online live trading platform. The advertised information on IG's website (IG 2021) is represented on the online demo trading platform. However, IG's online demo trading platform has a noticeable difference with the IG's live account. The minimum bet size (trade size) available in the live account is different from what has been advertised and what is shown on the IG online demo trading platform.

|  |  | Secondary Currency (columns) | | | | | |
|---|---|---|---|---|---|---|---|
|  |  | DKK | HUF | RUB | PLN | USD | ZAR |
| **Primary Currency (rows)** | AUD/ |  |  |  |  | 1 (0.5) |  |
|  | EUR/ | 0.2 (1) | 0.5 (1) | 0.02 (1) | 0.5 (1) |  |  |
|  | GBP/ |  |  |  |  | 0.5 (1) |  |
|  | USD/ | 0.5 (1) | 0.5 (1) | 0.02 (1) | 0.5 (1) |  | 0.1 (1) |

Table 3: Examples of discrepancies in minimum bet size in pounds (£) between live and online demo trading platforms. Rows show the primary currency and columns show secondary currency, e.g., minimum bet size on online live trading platform for EUR/DKK is £0.2 whereas on the online demo trading platform is £1 or minimum bet size on online live trading platform for AUD/USD is £1 whereas on the online demo trading platform is £0.5.

The profit and loss discrepancy was recorded as low as 6% on the demo account and as high as 50% on the live account, identified in Section 3.1. Which means whilst trading on the online demo trading platform, if you win, you win 6% less than you should win, and if you lose, you lose 6% more than you should lose.

However, the online demo trading platform suffers from the same technical issues as the online live trading platform.

## 3.3 Evaluation of Trustpilot reviews

Table 4 illustrates the overall summary of downloaded reviews. Of 4603 reviews on Trustpilot between 2015 to 2021, there were 3195 English reviews which are only considered in this study and 2136 received an invitation from IG to provide feedback and 1854 of them are verified by IG. There were 1058 reviews which were initiated by IG's clients on their own willingness.

|  |  | Frequency (%) |
|---|---|---|
| Language | English | 3195 (69.4%) |
|  | Non-English | 1408 (30.6%) |
| Invited by IG | Invited | 2137 (66.8%) |
|  | Not invited | 1058 (33.2%) |
| Verified | Yes | 1855 (86.8%) |
|  | No | 282 (13.2%) |
| Has the client reviewed other companies than IG? | Reviewed only IG | 1323 (41.4%) |
|  | Reviewed other companies than IG | 1872 (58.6%) |

Table 4: Number of English reviews, reviews by invitation and verified reviews.

Table 5 is the breakdown of 1 to 5 ratings based on if clients were invited to leave feedback or not. There is a significant association between rating and review by invitation (p-value <2.2e-16). This means it is more common for clients to initiate on their own willingness 1 star reviews than 5 star reviews and that IG sends invitations to customers who are likely to leave 5 star reviews.

|  | Rating | | | | |
|---|---|---|---|---|---|
| Invited by IG | 1 | 2 | 3 | 4 | 5 |
| Not Invited | 623 (87.1%) | 68 (50.4%) | 45 (17.9%) | 70 (11.2%) | 252 (17.2%) |
| Invited | 92 (12.8%) | 67 (49.6%) | 206 (82.1%) | 557 (88.8%) | 1215 (82.8%) |
| Total | 715 (22.4%) | 135 (4.2%) | 251 (7.9%) | 627 (19.6%) | 1467 (54.9%) |

Table 5: Ratings grouped by whether clients were invited to leave a feedback.

### 3.4 Technical issues based on Trustpilot reviews

Figure 3 shows the frequency 1-5 star ratings between 2015 to 2021. It suggests the number of 5 star reviews have dramatically reduced by 51.9% from 397 in 2018 to 191 in 2021. However, the number of 1 star reviews considerably increased by 346.2% from 81 in 2018 to 277 in 2021.

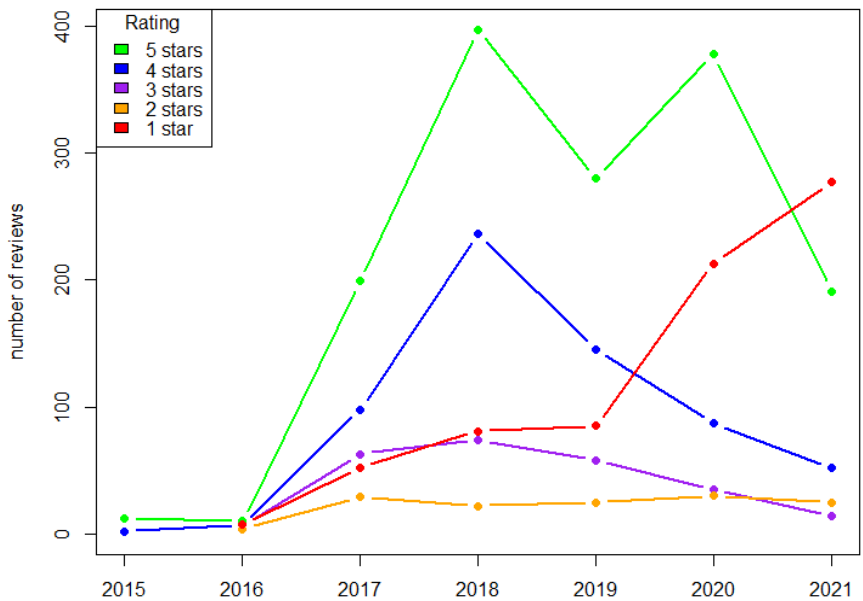

Figure 3: Number reviews broken down by the 1-5 star ratings between 2015 to 2021.

Figure 4 illustrates the daily count of 1 star IG's Trustpilot reviews reporting any platform technical issues between 2015 and 2021. It can be seen that these technical issues are still ongoing and not isolated to specific time periods. From 2020 more 1 star reviews were reported.

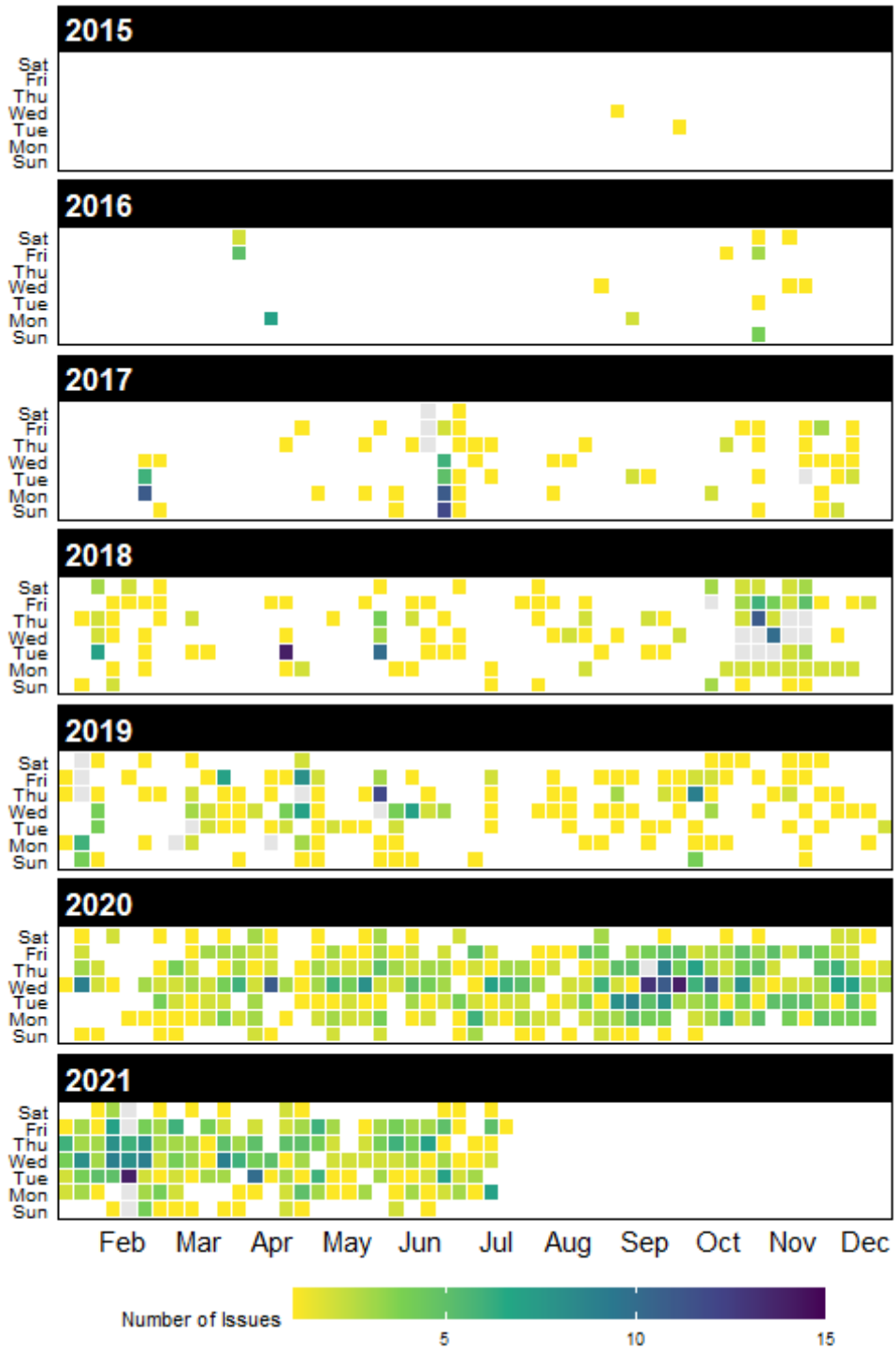

Figure 4: Timeline of 1 star reviews. A random occurrence of the 1 star reviews but persistence over six years (2015-2021) with a maximum number of 250 1 star reviews in one day issues reported on 15 July 2017.

Table 6 summarises major technical issues reported in Trustpilot reviews. The most frequent comment was related to an unstable online trading platform. The second most frequently reported issue was IG changing its rules unfairly without transparency. Stop limits[5] that were triggered too early and not being able to close open positions were the third and fourth most frequent issues, respectively.

Between May and June 2021 there were five clients who stated that their profit disappeared from their statements. The time period matches with the disappearance of 13 trades identified in Section 3.1.

In general, comparing Tables 2 and 6 confirms that the unstable trading platform, inability to close positions manually or using REST API are the most frequent technical issues.

| Identified Issue | Frequency of the issue | Financial loss/damages |
|---|---|---|
| Unstable online trading platform | 227 | 59,506 |
| IG's changed its rules unfairly without transparency | 148 | 265,877 |
| Stop limit triggered too early | 120 | 5,300 |
| Unable to close trades | 116 | 42,005 |
| Unable to withdraw funds | 82 | |
| Unable to open trades | 65 | 2560 |
| Limit order to take in triggered too early | 50 | |
| Not being able to log in | 46 | |
| Discrepancies between demo account and live account | 21 | |
| No open position appears on the online trading platform | 16 | |
| Incorrect statement | 7 | |
| Disappearance of profit | 5 | 23,501 |
| **Total** | **903** | **356,744** |

Table 6: Details of frequent issues reported by client and their frequency of occurrences based on IG's 1 star Trustpilot client reviews.

---

[5] "A stop limit is the maximum amount of money that a trader is prepared to lose on a particular trade" (Hayes 2021).

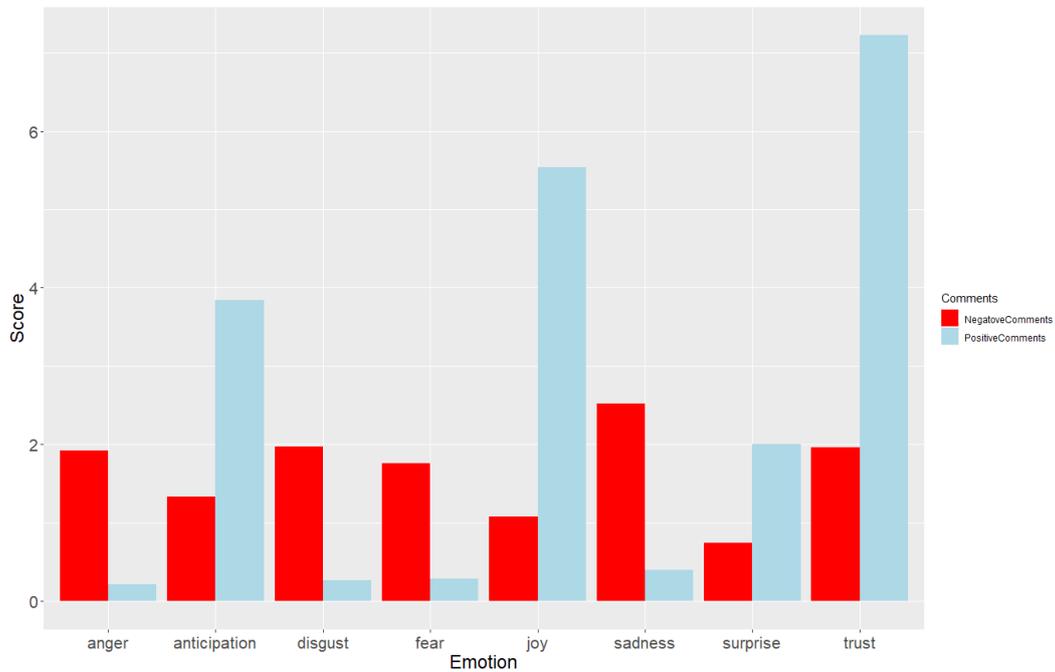
Figure 5: Commonly used description of traders' experience after contacting IG.

## 3.5 Customer service response to IG's technical issues

When clients face any technical issues, it is obvious to contact their customer support for help. It is expected that any advice given should be clear and concise.

From 1467 customers who left 5 star reviews for IG, 82% of them used phrases that were for new customers of IG who found it "easy to open and set up an account" with IG. They commented that the "platform is easy to use for beginners".

However, of 699 customers who left 1 star reviews with comments, 95.9% of them used phrases that were from existing customers who found that IG had the "worst customer service", "unhelpful" and "shocking''.

Figure 5 summarises how customers described their experience with IG customer service based on Trustpilot reviews. The customers with 5 star reviews expressed joy and trust when opening their new account with IG. However, when customers contacted IG with a 1 star review about their issues, they felt lack of trust, anger and sadness.

For this study, IG customer service was contacted to help resolve issues presented in Table 2. Below we summarised the experience from communication with IG between 17 May 2021 and 12 July 2021 (Table 7).

In general, IG was not informative with regards to their online trading platform technical issues. They initially refused to confirm that there was an issue, however the trader was encouraged to use the online trading platform to check if IG's problems persisted. On 5 July 2021, IG finally confirmed that there was an issue of trading latency in May 2021. However, IG refused to confirm whether these issues have been resolved or not. Instead, IG asked for an update whether or not these technical issues still existed.

|   | **Communication with the broker** | **Authors Comments** |
|---|---|---|
| 1 | Broker was aware of a confirmed serious issue and they failed to inform the trader. | Between 17 May 2021 and 22 June, the broker did not either confirm or deny that there was a technical issue with its online trading platform. On 5 July 2021, the broker confirmed that there was a confirmed issue with its online trading platform. During the period of this study the broker refused to provide any updates on their confirmed technical issue. |
| 2 | Broker whilst investigating knew there were confirmed issues that can affect some trades. | |
| 3 | Broker had to inform its clients of any updates of existing issues that could affect clients' trading. | |
| 4 | All clients have the right to know if there are issues on Broker's online trading platform before they trade. | It is the broker's responsibility to inform their clients on ongoing technical issues on its online trading platform. Then, it will be clients' choice to accept further risk and trade on the broker's online trading platform. The broker has taken this right away from its clients. |
| 5 | If I was aware of these issues I would not have traded. | |
| 6 | Of 168 trades between 4 May 2021 to 22 June 2021, all had one or more of Broker's issues (see Table 2). | |
| 7 | As soon as the trader established there was a problem with Broker's online trading platform the trader stopped trading. | Without providing any updates on technical issues in Table 2, the broker requested its client to inform them if the trader was still experiencing the same issues. Once the trader informed the broker that he is not willing to trade on broker's unstable online trading platform the broker informed the trader they will close their investigation if they do not receive any information from the client. |
| 8 | Four times Broker asked the trader for information which absolutely required me to be in trades. | |
| 9 | Twice Broker said if they didn't get information from me, they would close investigations. | |
| 10 | Serious issues with Broker's system could be simply verified by looking at the log files | Every online trading platform keeps the log record of traders' activities and interaction with their online trading platform. Any technical issues, error messages and unsuccessful attempts to login, place order or close a trading position available through these log files. However, the broker refused to refer to these log files and ignored the request of the trader to access his incomplete log files until 12 July 2021. |
| 11 | All issues in Table 2 came from broker's REST API responses. | |
| 12 | The conclusive evidence of all broker's client trading activity is in the log files. | |
| 13 | Log files are traders' personal information which they have the right to access on request. | Log files can be considered as traders' personal information which the broker has to provide this information to the trader under the Data Protection Act 2018 (GOV.UK 2018). |
| 14 | As a client I have the right to request my log files, this request has been ignored | |
| 15 | Broker concluded an investigation without providing clients with any log files or any evidence. | To date, the broker refuses to provide the comprehensive log files as requested by the trader under the Data Protection Act 2018 (GOV.UK 2018). |

Table 7: Summary of communication with IG.

# 4   Discussion

IG is an example of brokers who carry badges of approval from various regulatory bodies whilst intelligently manipulating its online trading platform. If relying solely on statements provided by the broker, it is virtually impossible to identify these manipulations.

IG's lack of transparency and unwillingness to address or inform its clients about these online trading platform technical issues brings us to the conclusion that IG is using these platform technical instabilities to control retail traders' profit and loss.

IG is one of the world's leading online trading providers, offering its services across seventeen countries. This means IG is licensed under various regulatory bodies.

In the United Kingdom, out of 496 complaints in total made against spread betting brokers to Financial Ombudsman Services (FOS), 168 (33.8%) of those were upheld. Out of that 496 of complaints, 165 (33.2%) of those complaints were made against IG, of which only 20 (12%) of those complaints were upheld (Financial Ombudsman Services 2021).

IG has shown high instability on its online trading platform which has resulted in significant losses for traders, especially over the past six years. These include not being able to close profitable trades, closing trades with delay, disappearance of trades, disappearance of profit from clients statements, profit and loss discrepancies, stop loss not being triggered, stop loss or limit order triggered too early. These are the contributing factors for retail traders' losses in addition to the 82% risk of losing money identified by the FCA.

These technical issues can result in one retail trader losing from a few pounds to thousands of pounds. However, IG having over 185,000 retail investors, (IG Group 2020) this can be a significant misappropriation of profit and manipulation by IG through its online trading platform.

Not having log files readily available for retail traders is a prime example of non-transparency. Access to log files is the only way to identify any platform technical issues because log files contain all retail trader's detailed trading activities and online trading platform interactions. Therefore, it is essential for any retail trader to have access to their log files to assess any suspected profit or loss manipulation by the broker. Although, these log activity files can be altered, deleted or doctored by the broker.

The regulatory bodies rarely consider broker's manipulation of profit and loss through online trading platform technical issues as misconduct or crimes (Bloch 2020). Even if they did, they treat these acts with a token verbal guilty verdicts and slap on the wrists fines and small compensation payouts for retail traders. This gives no real reason for such broker activities to ever stop such manipulations.

Retail traders' complaints should not be considered as just an emotional reaction of retail traders to losing money. Although as previously mentioned, for retail traders unexpected financial loss due to technical online trading platform issues can be as traumatising as would the sudden loss of a loved one, being harassed, mugged or robbed.

# 5 Conclusion

In this paper we studied how retail traders' profit and losses can be affected by spread betting broker's online trading platform technical issues. We introduced the idea that technical issues can be considered as online trading platform manipulations to control traders' profit and loss.

As an example, we identified trading platform technical issues of one of the world's leading online trading providers, IG. We showed that such platform technical issues can result in significant financial losses for retail traders.

We compared and identified issues in this study with issues highlighted by IG's 1 star Trustpilot reviews. Our study revealed that IG's clients have been suffering from these technical issues since at least 2015 and are getting worse. We found that IG's customer services with regards to these identified issues were dishonest, unhelpful and untransparent.

From IG's Trustpilot 1 star reviews, it was revealed that some traders record their evidence of trades by taking screenshots of their open and closed positions to protect their assets. However, this is labour intensive, inaccurate and an impossible task to do in regard to checking the details of a trading account. To overcome this problem, we used IG's REST API responses as the proof of all trading activities and manipulation of profit and loss by IG.

However, unless the trader is an advanced level programmer it would be very difficult to use REST API. Also, not all brokers may provide REST API trading. Therefore, it is essential and it should be made mandatory for retail traders to have immediate access to their entire log trading activity files.

To conclude, it is very easy for a broker to manipulate their online trading platform by introducing technical issues to control traders' profit and losses. Therefore, regulatory bodies such as the FCA should take these technical issues more seriously and not rely on brokers' internal investigations, because under any other situations or circumstances, these platform manipulations of retail traders' profit and loss would be considered as crimes, fraud and connivingly misappropriation of funds.